\documentclass[prb]{revtex4}
\usepackage{graphicx}

\begin{document}

\author{Alejandro Cabo$^{1,2}$, Francisco Claro$^{3}$ and Norman H. March$^{1,4}$}
\title{Melting curves and other phase transitions in a two dimensional electron
assembly in a transverse magnetic field as a function of the Landau level
filling factor }

\affiliation{$^{1}$Abdus Salam International Center for
Theoretical Physics, Trieste , Italy}

\affiliation{$^{2}$ Grupo de F\'{\i }sica Te\'{o}rica, Instituto
de Cibern\'{e}tica y F\'{\i }sica,  La Habana, Cuba. }

\affiliation{$^{3}$ Facultad de F\'{\i }sica, Pontificia
Universidad Cat\'{o}lica de Chile, Santiago de Chile, Chile,}

\affiliation{$^{4}$ Department of Physics, University of Antuerp,
Antuerp, Belgium and Oxford University, Oxford, England.}

\begin{abstract}
\noindent Besides Laughlin/composite Fermion liquid, and Wigner solid, there
have been proposals for Hall crystals and condensed phases of skyrmions for
inclusion in the QHE systems phase diagram. Some results on \ Hall crystals
are reported, which suggest \ that at $\nu =1/3,$ such crystals can be the
ground state of the 2DEG in a magnetic field. At $\nu =1$ and $2$, Chen $et\
al$ experimentally established that the Wigner solid is the ground state.
An explanation is also given here for the coexistence in these experiments of 
an integral QHE plateau and the presence of the  Wigner crystal phase near $nu=1$.
  Further, Mandal $et\ al$
showed that at $\nu =1/7,$ and most probably \ at $\nu =1/9,$ an
incompressible liquid state should be the ground state. Thus, at least three
non-entrant solid phases \ appear to exist: $0<\nu \leq 1/9, 1/9\leq \nu
\leq 1/7$ and, less certainly, \ $1/7\leq \nu \leq 1/5.$ In addition, points
of solid nature \ are present at $\nu =1$ and \ $2$ \ (Wigner solid) and
possibly at $\nu =1/3$ (Hall crystal). \

\bigskip

\noindent PACS numbers: 73.43.Cd, 73.43.-f

\bigskip
\end{abstract}

\maketitle

\affiliation{$^{1}$Abdus Salam International Center for
Theoretical Physics, Trieste , Italy}

\affiliation{$^{2}$ Grupo de F\'{\i }sica Te\'{o}rica, Instituto de
Cibern\'{e}tica y F\'{\i }sica,  La Habana, Cuba. }

\affiliation{$^{3}$ Facultad de F\'{\i }sica, Pontificia Universidad
Cat\'{o}lica de Chile, Santiago, Chile}

\affiliation{$^{4}$ Department of Physics, University of Antuerp, Antuerp,
Belgium and Oxford University, Oxford, England.}

\bigskip \ The boudaries between different phases of the 2DEG at high
magnetic fields has been the subject of much interest recently.A schematic
diagram of the melting curve characterizing the equilibrium \ between a
Wigner solid and a Laughlin liquid \ was proposed more than a decade ago by
Buhmann $at$ $al.$ \cite{buhm,price} This was followed by a thermodynamical
discussion based on a first order\ phase transition \ by Lea, March and Sung
\cite{lea1}. These authors \ wrote for the slope of the melting curve $%
T_{m}(\nu ),$ with $\nu $ the Landau level filling factor,

\begin{equation}
\frac{\partial }{\partial \nu }\ T_{m}(\nu )=\frac{B}{\nu }\frac{\Delta M}{%
\Delta S}.  \label{eq1}
\end{equation}
Here $B$ is the strength of the transverse magnetic field, while\ $\Delta M$
and $\Delta S$ are the changes in the magnetization $M$ and entropy $S$
across the melting line. To be precise, with $"\ s\ "\ $solid, and $"\ l\ "$
liquid :
\[
\Delta M=M_{s}-M_{l}.
\]
and
\[
\Delta S=S_{s}-S_{l}.
\]
Lea, March and \ Sung \cite{lea} subsequently applied microscopic theory,
for both anyon and composite Fermion models, to demonstrate that the main
features of the Buhmann $et$ $al$ diagram could be explained by invoking the
magnetization, of the de Haas-van Alphen character, of the Laughlin/CF
liquid \ \cite{koh,mar}.

Key additions to this field are the very recent studies of Mandal, Peterson
and Jain \cite{mandal} who presented \ Monte Carlo results for $\nu =1/7,1/9$
and $1/11.$ They then write that `the principal conclusion of this work is
that, for the model considered, the ground state is a liquid at $\nu =1/7$'.
This is in agreement with the Buhmann $et$ $al$ schematic diagram. \ The
second addition, also very recent, is the experimental study of Chen $et$ $%
al.$\cite{chen} on microwave resonance measurements which establishes that a
pinned crystal is the ground-state at both $\nu =1$ and $2$. Some
explanations about their experiment are offered below. These two new
contributions will here be embodied\ in a proposal to refine and extend
somewhat \ the phase diagram of Buhmann $et$ $al$. In the course of such
refinement, we shall want to consider further phases besides the Wigner
solid: \ a `Hall'crystal (HC) and a `Skyrme' crystal (SC).
\begin{figure}[tbp]
\includegraphics[width=4.in]{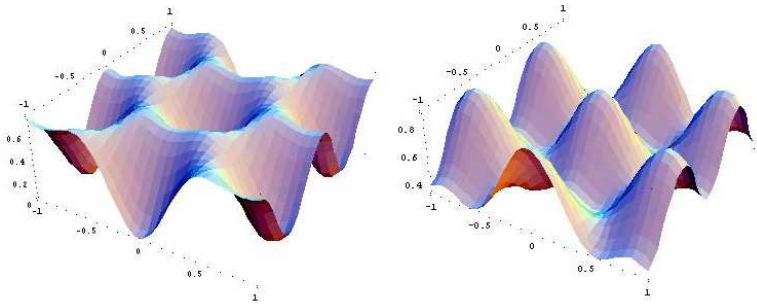}
\caption{Charge density wave profile of the Hartree-Fock approximation for
the electron distribution in the Hall crystal (left). For comparison, the
Wigner crystal density is shown in the right (after Ref. 9)}
\label{fig:den1}
\end{figure}
Beginning with the HC state\cite{tesa}, two of us (A.C and F.C.) \cite
{cabcla} have focussed on the filling factor $\nu =1/3.$ Their starting
point was to build on the charge density wave states (CDW)\ found \ in early
work by F.C.\cite{claro1,claro2} using the Hartree-Fock (HF) approximation.
As $\nu $ is varied, the energy per particle of such states presents cusps
at all fillings with odd denominator revealing a gap, while they behave as
metals at all even denominators. Cabo and Claro\cite{cabcla} argued that in
addition the HC shows strong cohesive energy determined by cooperative rings
of exchange effects. The second order correlation correction turned out
sufficiently strong \ or the energy of such filling factors to go below that
of the Yoshioka-Lee HF\ approximation \cite{yosh} for the Wigner solid (WS).
The source of the intensity of these cohesive correlations is here
illustrated in Fig. \ref{fig:den1}, where the form of the electron density
of the HC state is shown in comparison with the one corresponding to the
Yoshioka-Lee state.

As can be observed, the channel like regions connecting the density maxima
imply that the Wannier like localized states of the problem should have
appreciable overlap. This is in contrast with the structure of the density
in the Yoshioka-Lee CDW that shows rather isolated gaussian-shaped peaks
with no channels between them. Therefore, the contributions of the
cooperative rings of exchange effects can be expected to produce substantial
cohesive effects. To be fully operative they require long paths and will not
be properly accounted for in numerical calculations involving small samples
and very few electrons. Although a precise evaluation of the total energy is
needed for a definitive conclusion to be extracted, at the present state of
understanding the above characteristics of these HC states lead us to
propose them here as feasible candidates for the ground state at $\nu =1/3$.
The investigation of this question will be considered in detail elsewhere.

\begin{figure}[tbp]
\includegraphics[width=4.0in]{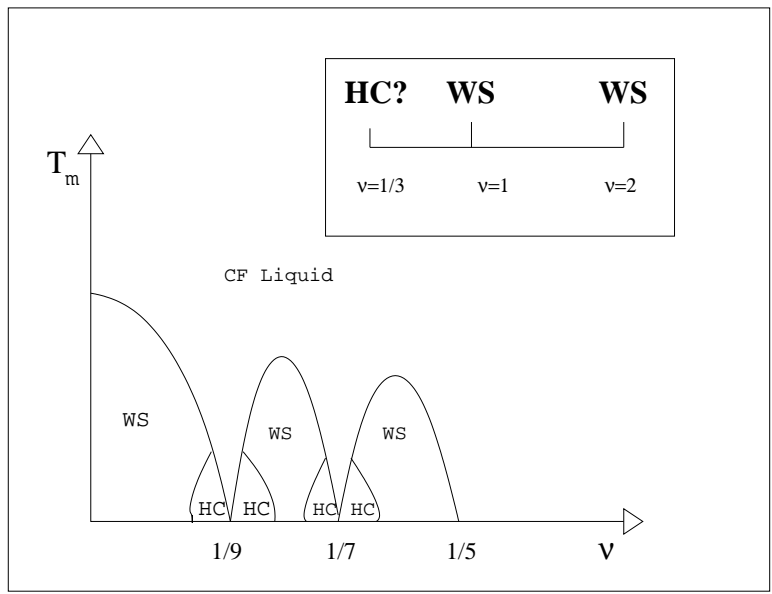}
\caption{Phase boundaries in the $T$ vs $\nu $ plane. WS is the Wigner Solid
phase, HC the Hall crystal, and CF indicates the composite fermion liquid.
The inset refers to the segment [1/3,2] of the filling factor}
\label{march}
\end{figure}

Besides the above implication that the HC is a strong candidate for the
ground state at $\nu =1/3,$ we have attempted in Fig. \ref{march} to refine
and extend the proposed phase diagram of Buhmann $et$ $al$ \cite{buhm} in
the light of these findings. We show $0<\nu \leq 1/3$ in the main panel and $%
\nu $ from $1/3$ to $2$ in the inset. Although still very schematic, we
think the HC phase may come into its own\ in the three `solid' reentrant \
phases shown, and have then sketched further the phase boundaries between
the HC and WS states. In the inset, it is relevant to add the ground states:
both WS \cite{chen} near $\nu =1$ and $2$. It should be remarked that an
earlier proposal was made for n SC\ around $\nu =1$\cite{brey,green,cote}.
While we in no way rule out the possible appearance of an SC\ phase
elsewhere in the schematic phase diagram of Fig. \ref{march}, Chen $et$ $al$
argue conclusively that, since Skyrmions do not exist at $\nu =2$ and their
microwave resonances show major similarities\ around $\nu =1$ and $\nu =2$,
the Skyrmion crystal is not the explanation for their data near $\nu =1.$
They present conclusive evidence that the microwave resonance is caused by
the pinning mode of a crystalline phase.

In connection with $\nu =1$ and $2$ we would like to also include here some
comments related with the detection of WS like states in the work of Chen $et
$ $al.$ \cite{chen} They find that the magnetic field region in which the
microwave absorption shows the peaks are well inside the plateaus.\cite{tsui}
\ Then, the question about the compatibility of the crystalline nature of
the ground state and the quantized value of the Hall conductivity emerges.
We just want to point out in this note that the occurrence of such an effect
is coherent \ with the results of the work of Cabo and Martinez.\cite{cabdan}
It is there argued that the linear response of a 2DEG \ in the integral
quantum Hall regime can be roughly described as `forcing' the impurities \
to act as effective `charge reservoirs', receiving or releasing the exact
amount of electrons to ensure the local integral filling condition in large
sample areas. \ Therefore, the following picture seems a plausible way to
explain the findings of Chen $et$ $al.$\cite{chen} First, when the filling
factor is near the values $\nu =1,2$ for a very clean sample, it can be
expected that the 'charge reservoirs' accumulating the excess or defect
charges are the small number of imperfections present in the clean sample. \
However, when $B$ deviates more, $ie$. at $\nu =$ 8/9, (10/9), it could
happen that the limited quantity of localization centers in the high quality
sample employed are saturated to their capacity. \ Therefore, a next
imaginable step for the system to continue localizing defect (excess)
charges is to start situating them in pinned crystalline regions which could
be viewed as a system of dynamically generated localization centers.

\ Returning to the proposed phase diagram in Fig. \ref{march}, we notice
that near $\nu =1/q$ the form shown suggests melting of the HC, but to a
solid-solid phase transition further away. If our proposal proves to be
useful, the WS is always the high temperature phase of those solid-solid
transitions since correlatiosn favor the HC state. Due to the thermodynamic
result (\ref{eq1}), which of course applies everywhere to the phase
boundaries shown in the figure, assuming that they are all representing
first-order phase transitions the magnetization $M$ and the entropy $S$ of
the HC are subjects well worth further investigation. For the CF liquid, $%
M_{l}$ and $S_{l}$ were treated at some length more than a decade ago by
Lea, March and Sung \cite{lea}.

\section{Acknowledgments}

A.C. acknowledges the kind hospitality of the AS ICTP\ during the visit to
the Center allowing this collaborative work. The support of \ FONDECYT
grants 1020829 and 7020829 for the activity of \ F.C. is very much
recognized. N.H.M wishes to acknowledge that his contribution to this study
was made during a visit to the Condensed Matter Group at the AS ICTP,
Trieste. He wishes to thank Prof. V.E. Kravtsov for the stimulating
atmosphere provided and for generous hospitality.

\end{document}